\documentstyle[11pt,newpasp,twoside,epsf]{article}
\def\edcomment#1{\iffalse\marginpar{\raggedright\sl#1\/}\else\relax\fi}
\reversemarginpar

\begin{document}
\title{Resolving Sirius-like binaries with the Hubble Space 
Telescope}
 \author{Matt Burleigh and Martin Barstow}
\affil{Dept. of Physics and Astronomy, University of Leicester, Leicester 
LE1 7RH, UK}
\author{Howard E. Bond}
\affil{Space Telescope Science Institute, 3700 San Martin Drive, Baltimore, 
MD 21218, USA}
\author{Jay Holberg}
\affil{Lunar and Planetary Laboratory, Gould-Simpson Building, University of 
Arizona, Tucson, AZ 85721, USA}

\begin{abstract}
We have imaged seventeen recently discovered 
Sirius-like binary systems with {\it HST/WFPC2} and 
resolved the white dwarf secondary in eight cases. Most of the implied 
orbital periods are of order several hundred years, but in three cases 
(56 Per, $\zeta$ Cygni and RE~J1925$-$566) the periods are 
short enough that 
it may be possible to detect orbital motion within a few years. It will 
then be possible to 
derive dynamically determined masses for the white dwarfs, 
and potentially these stars could be used as 
stringent tests of the mass-radius 
relation and initial-final mass relation. 
 
\end{abstract}

\section{Introduction}

Although universally accepted, the white dwarf mass-radius relation 
(Chandrasekhar 1931; Hamada \& Salpeter 1961; Wood 1992) 
has resisted 
in-depth examination through direct observations for $\approx$70 years. The 
difficulty arises mainly from obtaining accurate, model-independent 
measurements of the masses and radii of known white dwarfs. A combination of 
spectroscopically derived temperatures and gravities, and accurate parallaxes 
can provide the independent means of determining $M_{\rm WD}$ and 
$R_{\rm WD}$, and therefore a new route for examining evolutionary models. The 
white dwarfs observed by {\it Hipparcos} (Vauclair et~al. 1997; Provencal 
et~al. 1998) were entirely consistent with expectations, although they were 
restricted mainly to a narrow range around $0.6M_\odot$. 
Furthermore, even with these new parallaxes the overall uncertainties are 
still too large to provide a really thorough examination of the differences 
between theoretical models which assume a variety of core and envelope 
compositions.

White dwarfs in resolved binary systems can provide the most stringent tests 
of evolutionary models, since the mass can be determined from 
the orbital and physical elements of the system. In practise, though, few 
such systems are available to be studied in sufficient detail (they are: 
Sirius, Procyon, 40 Eri and Stein 2051). However, as reported 
at previous workshops (e.g.~Burleigh 1999) $>20$ new Sirius-like binaries 
have been identified through the {\it ROSAT WFC} and {\it EUVE} all-sky 
surveys. Each system consists of a normal or subgiant star plus a hot white 
dwarf companion, which is responsible for the soft X-ray and EUV flux 
detected by {\it ROSAT} and {\it EUVE}. Crucially, in most cases, the bright 
primary has an accurate {\it Hipparcos} parallax, yielding a precise distance 
for the white dwarf. Thus this new sample of Sirius-like binaries presents a 
golden opportunity of extending the sample of well-studied  
white dwarf binary systems. Our first aim, therefore, is to identify those 
systems that can be resolved and have their orbits measured within an 
acceptable timescale. Unfortunately, none of these new binaries can be 
resolved from the ground, due to the huge difference in brightness between 
the components ($>5$ mags.), although we know from radial velocity studies 
that most of these systems are wide with periods $>$few years (e.g.~Vennes, 
Christian, \& Thorstensen 1998).

The {\it Hubble Space Telescope} provides an answer to this problem. 
Unencombered by the Earth's atmosphere, it allows imaging in the UV where 
the brightness of the hot white dwarf and its companion are similar. It also 
delivers a diffraction limited resolution of $\approx0.05\arcsec$, making it 
possible to resolve binary components with separations as small as 0.1\arcsec. 
We report here the first results of a snapshot survey with {\it HST/WFPC2} to 
image these new Sirius-like binary systems.

\section{HST/WFPC2 imaging}

All the observations have been conducted with the Planetary Camera CCD chip of 
the {\it WFPC2} instrument aboard {\it HST}. A single UV filter was chosen for 
each system to give approximately equal fluxes between the two components. 
Exposure times ranged from 8 to 500 seconds, calculated to give the maximum 
possible signal-to-noise without saturating the chip. Seventeen systems 
have been imaged thus far, and we have resolved the components in eight 
cases (Figures 1 \& 2). A further $\sim$ten systems await scheduling. 

For the resolved binaries we have  
calculated the angular separation and, using the known distance to each 
system, converted this to a physical separation perpendicular to the 
line of sight (Table 1). 
Note that these values most likely represent lower limits on the true 
separation, which will depend on the orbital inclination and current phase. 
From Kepler's third law it is then possible to calculate a lower limit on the 
orbital period, although note we have assumed a mass of $0.6M_\odot$ for 
each white dwarf (the primary masses are estimated from their spectral type).

For the unresolved systems, we can estimate the projected physical 
separations and orbital periods, assuming a minimum measurable separation on 
the CCD chip of 0.083\arcsec (Table 2). These results should of course 
be treated with a certain amount of caution, since such apparantly small 
separations may be observed if the components are close to conjunction 
or opposition. The true separation and inferred orbital period could 
be much larger than indicated here.

Of the resolved systems, three immediately stand out. 56 Per, which is in fact 
a quadruple system (the known visual companion is also resolved into two 
components), has a nominal period of $\approx$47 years, similar to 
Sirius itself. RE~J1925$-$566 may have an orbital period as 
short as $\approx$70 years. In both these cases it may be possible to 
detect orbital motion within a few years. In addition, we have resolved 
the Barium giant $+$ white dwarf system $\zeta$~Cygni, which has a 
spectroscopic period of 6489$\pm$31 days (i.e. $\sim18$~years). The binary 
separation ($0.11\arcsec$) is consistent with this being the orbital period 
of the white dwarf. 

Among the unresolved systems, it is no surprise that the white dwarfs were 
not resolved in HR8210 (IK Peg) or HR1608 (63 Eri), since these systems have 
known binary periods of 21.7 and 903 days respectively. Our failure to resolve 
these white dwarfs increases the likelihood that the degenerate star is 
responsible for the radial velocity variations seen in the primary. 
Conversely, we have confirmed that the white dwarf in 14 Aur C is a wide 
member of a triple system, and the 2.99 day period seen in this system is 
most probably due to an M-dwarf companion (Holberg et~al. 2000). 

\begin{table}
\caption{Estimated separations and orbital periods for the resolved systems}
\begin{tabular}{llccc}
\tableline
System & & Distance & Separation & P$_{(nominal)}$ \\
       & & (pc)  & (arcsec) & (years) \\
\tableline
HD2133$^\star$ & F7\,V$+$DA & 140 & 0.602 & 590 \\
HD27483$^\star$ & F6\,V$+$DA & 46 & 1.276 & 260 \\
14 Aur C$^\star$ & F4\,V$+$DA & 82 & 2.006 & 1307\\
RE J1925-566 & G7\,V$+$DA & 110 & 0.217 & 95 \\
HD223816 & F8\,V$+$DA & 92 & 0.574 & 290 \\
56 Per$^\star$ & F4\,V$+$DA & 42 & 0.390 & 47 \\
MS 0354.6$-$3650 & G2\,V$+$DA & 400 & 0.992 & 6200 \\
$\zeta$ Cyg$^\star$ & G8\,IIIp$+$DA & 46 & 0.11 & 18 \\
\tableline
\tableline
$^\star$ Distance from Hipparcos 
\end{tabular}
\end{table}

\begin{table}
\caption{Projected separation limits and implied maximum orbital periods for 
the unresolved systems}
\begin{tabular}{llcc}
\tableline
System & & Distance & P$_{(upper)}$ \\
       & & (pc)  & (years) \\
\tableline
BD$+$08$^\circ$102 & F7\,V$+$DA & 62 & 9.7 \\
HD15638$^\star$ & F6\,V$+$DA & 199  & 47 \\
HD18131$^\star$ & K0\,IV$+$DA & 104 & 19 \\
HR1608$^\star$ & K0\,IV$+$DA & 55 & 7.3 \\
HR3643$^\star$ & F7\,II$+$DA & 139 & 21 \\
HR8210$^\star$ & A8\,Vm$+$DA & 46 & 4.6 \\
RE J1309$+$085 & F9V$+$? & 275 & 79 \\
BD$+$27$^\circ$1888 & F0\,V$+$DA & 250 & 62 \\
$\beta$ Crt & A1\,III$+$DA & 82 & 9.5 \\
\tableline
\tableline
$^\star$ Distance from Hipparcos
\end{tabular}
\end{table}

\begin{figure}
\plotone{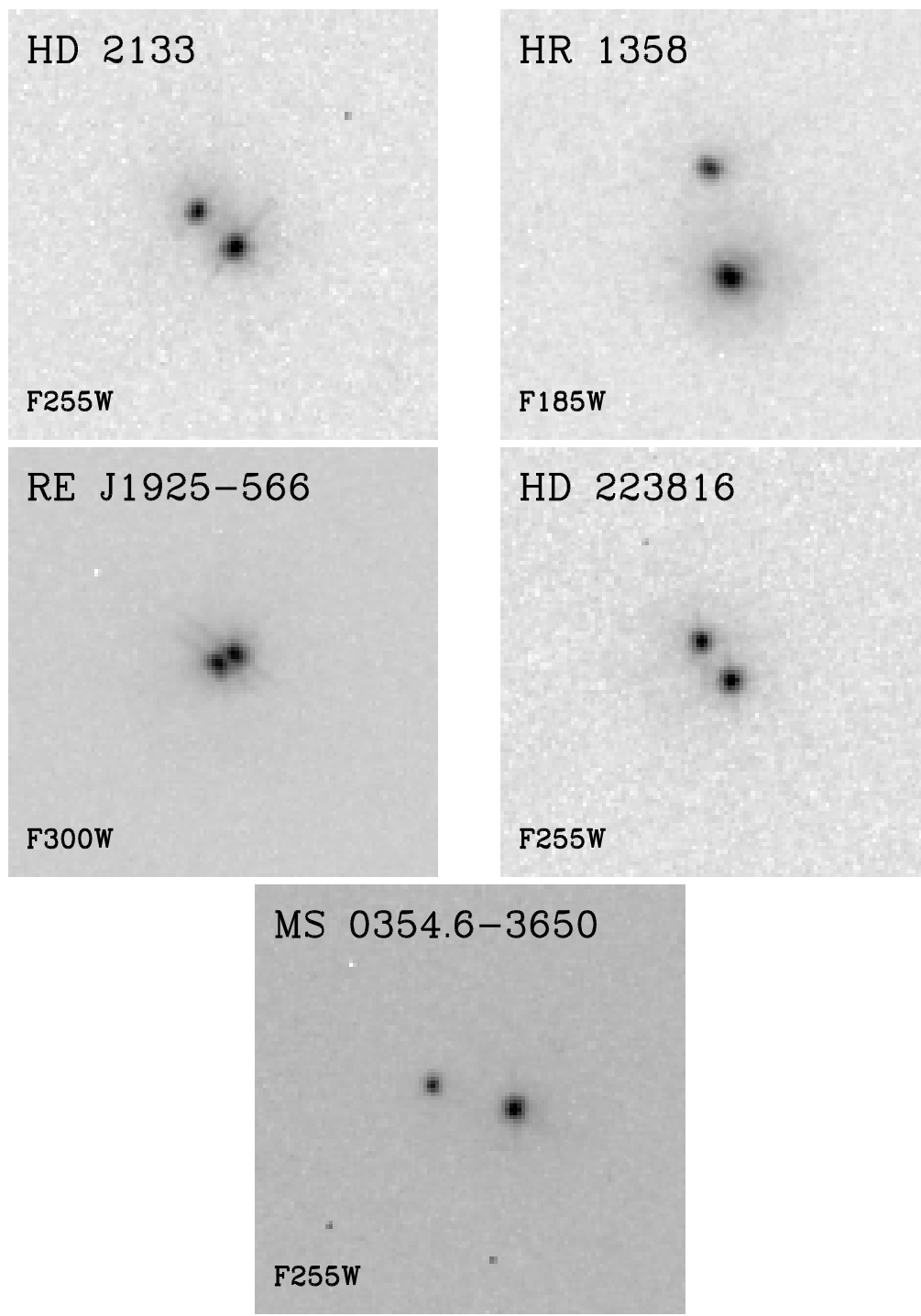}
\caption{WFPC2 snapshot images. Each frame is 
5$\arcsec$$\times$$5\arcsec$ in size, north is at the top and east on the 
left. The UV filter used is indicated on each image.}
\end{figure}

\begin{figure}
\plotone{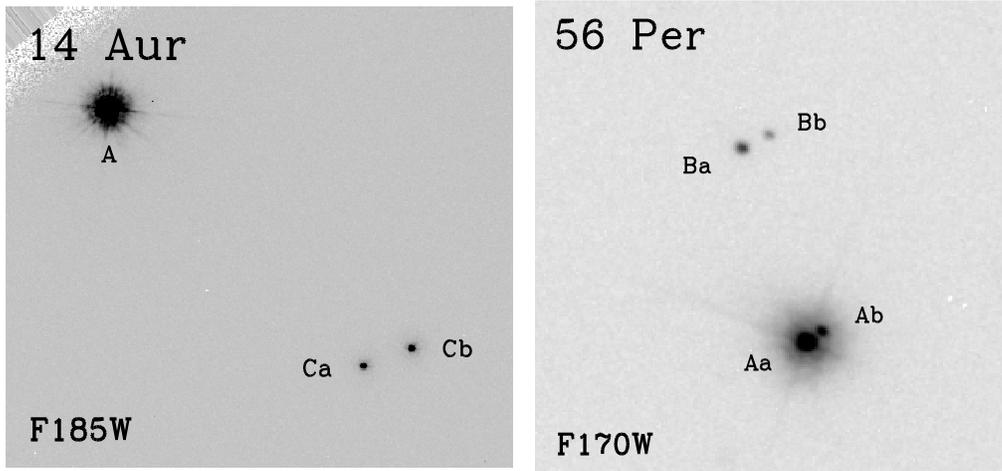}
\caption{Left: WFPC2 image of 14 Aur. The image is $20\arcsec$ wide. The 
C component is resolved into the object previously known from the 
ground, denoted Ca (itself an unresolved binary with a 2.99~day period), 
and the hot WD, denoted Cb. Right: 56 Per. This image is 
$10\arcsec \times 10\arcsec$. The Aa$-$Ab pair is the F4\,V primary plus 
WD companion. The known visual companion, 
56 Per B, is also resolved into two components.}
\end{figure}

\section{The next step: {\it FUSE} spectroscopy of the white dwarfs}

Surprisingly, we know very little about the white dwarfs in these systems. 
Since they cannot be resolved from the ground, it has thus far been 
impossible to obtain spectra of the H Balmer series in the optical region, 
from which the temperature ($T_{\rm eff}$) and gravity (log $g$) 
are usually derived. Instead, we have 
had to make do with any information we have been able to obtain from the 
single H Lyman $\alpha$ line and the shape of the continuum in {\it IUE} 
far-UV spectra. From these features 
alone it is impossible to unambiguously constrain $T_{\rm eff}$ and log $g$. 
However, now that the {\it Far Ultraviolet Spectroscopic Explorer (FUSE)} 
is fully operational, we can obtain spectra of the entire H Lyman series and 
use these absorption lines to derive $T_{\rm eff}$ and log $g$ (and hence 
mass and radius). We have a GI program with {\it FUSE} (PI Burleigh) 
to obtain these data for many 
of the new Sirius-like systems. Figure 3 shows the data from the first target 
to be observed, 14 Aur C. Modelling the Lyman lines with a pure-H atmosphere 
model, we obtain $T_{\rm eff}=41,000$K$-42,500$K and log $g=8.10-8.45$. Hence, 
$M_{\rm WD}=0.75-0.90M_\odot$ and $R_{\rm WD}=0.010-0.012R_\odot$, just about  
consistent with the lower limit on the 
{\it Hipparcos} distance estimate to the primary (74~pc). We stress that this 
represents work in progress, and we have some way to go before we will be 
fully confident in our modelling and understanding of the {\it FUSE} data. 

\begin{figure}
\plotfiddle{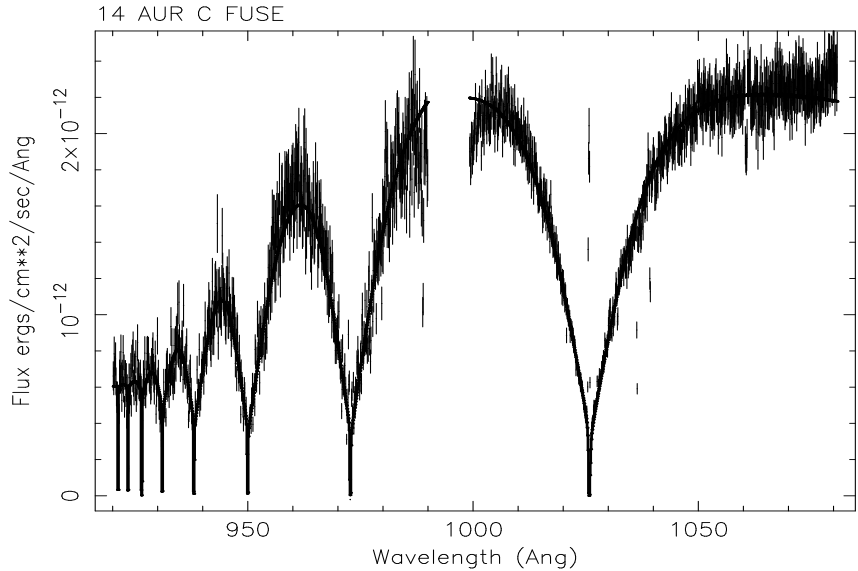}{6.5cm}{0}{67}{57}{-180}{5}
\caption{{\it FUSE} spectrum of the white dwarf companion to 14 Aur C. 
We have been able to model the data with a pure-H atmosphere 
$T_{\rm eff}=41,000$K$-42,500$K and log $g=8.10-8.45$.}
\end{figure}

\section{Propects from the {\it GALEX} far-UV all-sky survey and {\it SIM}} 

In 2002 NASA will launch {\it GALEX} (the {\it Galaxy Evolution Explorer}), 
a mission designed to survey the whole sky in the UV (Bianchi \& Martin 1998). 
This project is expected to detect $\approx3\times10^5$ hot white dwarfs 
(as well as $\approx10^6$ quasars!), and we calculate that $\approx30,000$ 
new Sirius-like binary systems will be identified. Obviously this is a huge 
dataset to follow-up, and so we have applied to the {\it Space Interferometry 
Mission (SIM)} (due for launch $\approx$2010) 
for Key Project Program status (PI: Holberg) 
to image, resolve and measure the orbits of 
many of the systems GALEX will identify.

\end{document}